\documentclass[longauth]{aa}
\usepackage{graphicx}
\usepackage[colorlinks]{hyperref}
\hypersetup{citecolor={blue}}
\usepackage{txfonts}
\usepackage{url}
\usepackage{float} % Required for [H] placement specifier

\begin{document}

   \title{Unveiling the 3D structure of the central molecular zone from stellar kinematics and photometry: The 50 and 20 km/s clouds}

\titlerunning{The 3D Structure of the CMZ: The 50 and 20 km/s Clouds}

\authorrunning{Nogueras-Lara et al.}
\author{
    Francisco Nogueras-Lara \inst{1,2} \fnmsep\thanks{E-mail: \url{fnogueras@iaa.es}}
    \and Ashley~T.~Barnes \inst{2}
    \and Jonathan D. Henshaw \inst{3,4}
    \and Karl Fiteni \inst{5,6}
    \and Yoshiaki Sofue \inst{7}
    \and Rainer Schödel \inst{1}
    \and Álvaro Martínez-Arranz \inst{1}
    \and Mattia C.~Sormani \inst{5}
    \and Jairo Armijos-Abenda\~no \inst{8}
    \and Laura Colzi \inst{9}
    \and Izaskun Jim\'enez-Serra \inst{9}
    \and Víctor M. Rivilla \inst{9}
    \and Pablo García \inst{10,11}
    \and Adam Ginsburg \inst{12}
    \and Yue Hu \inst{13}
    \and Ralf S.\ Klessen \inst{14,15,16,17}
    \and J.~M.~Diederik~Kruijssen \inst{18}
    \and Volker Tolls \inst{16}
    \and Alex Lazarian \inst{19}
    \and Dani R. Lipman \inst{20}
    \and Steven N. Longmore \inst{3,18}
    \and Xing Lu \inst{21,22}
    \and Sergio Mart\'{i}n \inst{23,24}
    \and Denise Riquelme-V\'asquez \inst{25}
    \and Jaime E. Pineda \inst{26}
    \and \'Alvaro S\'anchez-Monge \inst{27,28}
    \and Arianna Vasini \inst{5}
    \and Elisabeth A.C. Mills \inst{29}    
}

\institute{
    Instituto de Astrofísica de Andalucía (IAA-CSIC), Glorieta de la Astronomía s/n, E-18008 Granada, Spain
    \and European Southern Observatory, Karl-Schwarzschild-Strasse 2, 85748 Garching bei München, Germany
    \and Astrophysics Research Institute, Liverpool John Moores University, IC2, Liverpool Science Park, 146 Brownlow Hill, Liverpool L3 5RF, UK
    \and Max Planck Institute for Astronomy, Königstuhl 17, D-69117 Heidelberg, Germany
    \and Como Lake Centre for AstroPhysics (CLAP), DiSAT, Università dell'Insubria, via Valleggio 11, 22100 Como, Italy
    \and Institute of Space Sciences \& Astronomy, University of Malta, Msida MSD 2080, Malta
    \and Institute of Astronomy, The University of Tokyo, Mitaka, Tokyo, 181-0015, Japan
    \and Observatorio Astronómico de Quito, Observatorio Astronómico Nacional, Escuela Politécnica Nacional, 170403, Quito, Ecuador
    \and Centro de Astrobiología (CAB), CSIC-INTA, Carretera de Ajalvir km 4, 28850 Torrejón de Ardoz, Madrid, Spain
    \and Chinese Academy of Sciences South America Center for Astronomy, National Astronomical Observatories, CAS, Beijing 100101, China
    \and Instituto de Astronomía, Universidad Católica del Norte, Av. Angamos 0610, Antofagasta, Chile
    \and Department of Astronomy, University of Florida, P.O. Box 112055, Gainesville, FL 32611, USA
    \and Institute for Advanced Study, 1 Einstein Drive, Princeton, NJ 08540, USA
    \and Universität Heidelberg, Zentrum für Astronomie, Institut für Theoretische Astrophysik, Albert-Ueberle-Str. 2, 69120 Heidelberg, Germany
    \and Universität Heidelberg, Interdisziplinäres Zentrum für Wissenschaftliches Rechnen, Im Neuenheimer Feld 225, 69120 Heidelberg, Germany
    \and Harvard-Smithsonian Center for Astrophysics, 60 Garden Street, Cambridge, MA 02138, USA
    \and Elizabeth S. and Richard M. Cashin Fellow at the Radcliffe Institute for Advanced Study at Harvard University, 10 Garden Street, Cambridge, MA 02138, USA
    \and Cosmic Origins Of Life (COOL) Research DAO, \href{https://coolresearch.io}{https://coolresearch.io}
    \and Department of Astronomy, University of Wisconsin-Madison, Madison, WI 53706, USA
    \and University of Connecticut, Department of Physics, 196A Hillside Road, Unit 3046, Storrs, CT 06269-3046, USA
    \and Shanghai Astronomical Observatory, Chinese Academy of Sciences, 80 Nandan Road, Shanghai 200030, P.\ R.\ China
    \and State Key Laboratory of Radio Astronomy and Technology, A20 Datun Road, Chaoyang District, Beijing 100101, P.\ R.\ China
    \and 23European Southern Observatory, Alonso de Córdova 3107, Vitacura, Santiago 763-0355, Chile
    \and 24Joint ALMA Observatory, Alonso de Córdova 3107, Vitacura, Santiago 763-0355, Chile
    \and Departamento de Astronom\'ia, Universidad de La Serena, Ra\'ul Bitr\'an 1305, La Serena, Chile
    \and Max-Planck-Institut für extraterrestrische Physik, Gießenbachstraße 1, 85748 Garching bei München, Germany
    \and Institute of Space Sciences (ICE, CSIC), Campus UAB, Carrer de Can Magrans s/n, 08193 Bellaterra (Barcelona), Spain
    \and Institute of Space Studies of Catalonia (IEEC), 08860 Barcelona, Spain
    \and Department of Physics and Astronomy, University of Kansas, 1251 Wescoe Hall Drive, Lawrence, KS 66045, USA
}

   \date{}

% \abstract{}{}{}{}{} 
% 5 {} token are mandatory

  \abstract
   {The central molecular zone (CMZ), surrounding the Galactic centre, is the largest reservoir of dense molecular gas in the Galaxy. Despite its relative proximity, the 3D structure of the CMZ remains poorly constrained, primarily due to projection effects.}
   {We aim to constrain the line-of-sight location of two molecular clouds in the CMZ — the 50 and 20\,km/s clouds — and to investigate their possible physical connection using stellar kinematics and photometry. This study serves as a pilot for future applications across the full CMZ.}
   {We estimated the line-of-sight position of the clouds by analysing stellar kinematics, stellar densities, and stellar populations towards the cloud regions and a control field.}
   {We find an absence of westward moving stars in the cloud regions, which indicates that they lie on the near side of the CMZ. This interpretation is supported by the stellar density distributions. The similar behaviour observed in the two clouds, as well as in the region between them (the ridge), suggests that they are located at comparable distances and are physically linked. We also identified an intermediate-age stellar population (2–7\,Gyr) in both regions, consistent with that observed on the near side of the CMZ. We estimated the line-of-sight distances at which the clouds and the ridge become kinematically detectable (i.e. where the proper motion component parallel to the Galactic plane differs from that of the control field at the 3$\sigma$ level) by converting their measured proper motions parallel to the Galactic plane using a theoretical model of the stellar distribution. We find that the 50 and 20\,km/s clouds are located at $43\pm8$\,pc and $56\pm11$\,pc from Sgr\,A$^*$, respectively, and that the ridge lies at $56\pm11$\,pc; this supports the idea that the clouds are physically connected through the ridge.}
   {}

   \keywords{Galaxy: nucleus -- Galaxy: centre -- Galaxy: structure  -- Galaxy: stellar content -- infrared: stars -- proper motions -- dust, extinction
               }

   \maketitle%
%-------------------------------------------------------------------

\section{Introduction}

The central molecular zone (CMZ) is an accumulation of dense gas at the centre of our Galaxy \citep[e.g.][]{Henshaw:2022vl}. It extends over a radius of $\sim300$\,pc \citep{Morris:1996vn} and contains $3$–$5\times10^7$\,M$_\odot$ of dense clouds \citep{Dahmen:1998aa,Pierce-Price:2000aa}. This region arguably constitutes the most extreme star-forming environment in the Milky Way \citep{Lu:2019ab,Henshaw:2022vl,Hatchfield:2024aa} and is characterised by high temperatures \citep{Ginsburg:2016aa,Krieger:2017aa}, strong turbulence \citep{Federrath:2016vt,Kauffmann:2017aa,Henshaw:2019vh}, an intense magnetic field \citep[e.g.][]{Crutcher:1996aa,Pillai:2015aa,Hu:2022aa,Lu:2024aa}, and high confining pressures \citep{Yamauchi:1990aa,Spergel:1992aa,Muno:2004aa}.

\begin{figure*}
                 
                 \centering
   \includegraphics[width=0.9\linewidth]{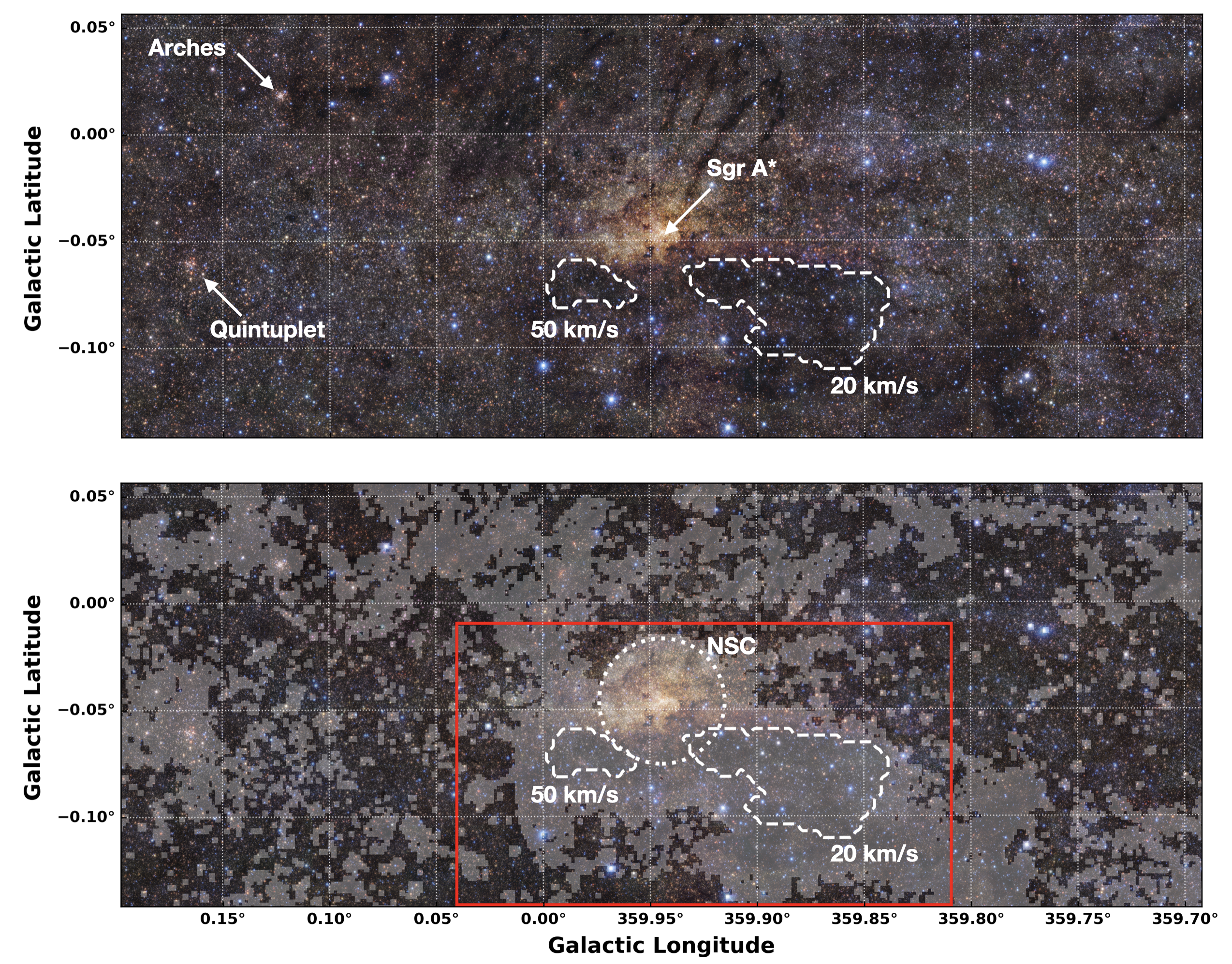}
   \caption{Upper panel: Central $\sim12'\times30'$ of the GALACTICNUCLEUS $JHK_s$  survey \citep{Nogueras-Lara:2019aa}, covering the 50 and 20\,km/s clouds (credit: ESO/Nogueras-Lara et al.). The positions of the 50 and 20\,km/s clouds are indicated, along with Sgr\,A$^*$ and the young stellar clusters Arches and Quintuplet. Lower panel: Control field (red box), where high-extinction regions (white shaded contours) are excluded, along with the 50 and 20\,km/s clouds (dashed white regions) and the nuclear star cluster (NSC; dotted white region), assuming an effective radius of $\sim4$\,pc \citep{Schodel:2014fk,gallego-cano2019}.}

\label{scheme}
\end{figure*}

The CMZ is shaped by gas funnelled through the Galactic bar \citep[e.g.][]{Sormani:2019aa} and likely gave rise to the nuclear stellar disc (NSD) — an old ($\gtrsim 8$\,Gyr), dense, and massive ($\sim10^9$\,M$_\odot$) structure at the Galactic centre. The NSD has a scale length of $\sim100$\,pc and a scale height of $\sim40$\,pc \citep[e.g.][]{gallego-cano2019, Sormani:2022wv}, and it partially overlaps with the CMZ \citep[e.g.][]{Launhardt:2002nx, Nogueras-Lara:2019ad}. The CMZ also hosts Sagittarius\,A$^*$ (Sgr\,A$^*$), the supermassive black hole at the centre of the Milky Way \citep[e.g.][]{Gravity-Collaboration:2018aa,Gravity-Collaboration:2020aa,Event-Horizon-Telescope-Collaboration:2022aa}.

Despite its relative proximity \citep[the Galactic centre is located at $\sim8.25$\,kpc; e.g.][]{Reid:2019aa,Gravity-Collaboration:2018aa,Gravity-Collaboration:2020aa}, deriving the CMZ structure is extremely challenging due to projection effects along the line of sight towards the Galactic centre \citep[e.g.][]{Battersby:2024aa}. Determining the 3D distribution of its gas is essential for: (1) understanding the formation of the NSD and its ongoing growth via newly formed stars; (2) pinpointing molecular clouds and star-forming regions to better comprehend their behaviour in the context of CMZ dynamics \citep{Henshaw:2022vl}; (3) tracing gas flows towards the central black hole to estimate mass transfer rates \citep{Schoedel:2023aa}; and (4) revealing the interactions between gas, stars, and the central black hole \citep{GRAVITY-Collaboration:2021aa}.

Efforts to determine the geometry of the CMZ have focused on converting the projected distribution of cloud positions and their line-of-sight velocities into a face-on morphological model \citep[see][and references therein]{Henshaw:2022vl}. While it is generally accepted that the CMZ gas follows an eccentric, ring-like, or toroidal structure, its precise geometry is still under debate. Three main models have been proposed: (a) a two-armed spiral \citep{Sofue:1995aa,Sawada:2004aa,Ridley:2017aa}, (b) closed elliptical orbits \citep{Molinari:2011fk,Walker:2025aa,Lipman:2025aa}, and (c) open streams \citep{Kruijssen:2015aa}. Several key molecular clouds are placed differently in existing models of the CMZ structure (Fig.\,3 in \citealt{Henshaw:2022vl}). In particular, the locations of the 50 and 20\,km/s clouds  \citep{Kauffmann:2017aa,Nogueras-Lara:2021uq,Henshaw:2022vl,Battersby:2024ab} remain unclear, with each model placing them at different positions along the line of sight. Determining their location is thus essential to help distinguish between competing models, and also to assess whether they interact with Sgr\,A$^*$ and supply material via gas inflow \citep[e.g.][]{Liu:2012aa, Lopez:2016aa, Tress:2020aa}.

In this work we combined stellar kinematics and photometry to investigate the line-of-sight location of the 50 and 20\,km/s clouds. This is possible thanks to the overlap between the NSD and the CMZ \citep[e.g.][]{Launhardt:2002nx}, which enables precise stellar photometry and astrometry even through high-extinction regions \citep[e.g.][]{Nogueras-Lara:2019aa,Libralato:2021td}. Our study builds on previous analyses of the Brick cloud \citep[e.g.][]{Nogueras-Lara:2021uq,Martinez-Arranz:2022uf} and represents a step forward since we are able to estimate line-of-sight distances. This work serves as a pilot study for our methodology, which will later be applied to the full CMZ to reconstruct its 3D morphology and assess which of the existing models — if any — best describes its actual structure.

\section{Data}
\label{data}

Figure\,\ref{scheme} shows the central $\sim12' \times 31'$ region ($\sim$\,28\,pc $\times$ 72\,pc, at the Galactic centre distance) of the NSD. This area includes the 50 and 20\,km/s clouds, whose line-of-sight positions we aimed to estimate. To identify stars associated with these clouds, we used the H$_2$ column density map from \citet{Battersby:2024aa,Battersby:2024ab} and defined the regions by selecting contours of $N(\mathrm{H}_2) = 1.1\times10^{23}\,\mathrm{cm}^{-2}$ that outline the clouds as shown in Fig.\,2 of \citet{Battersby:2024aa} and in our Fig.\,\ref{scheme}.

We also defined a control field within the same area, at similar Galactic longitude and latitude as the target clouds, avoiding regions with high extinction that are not representative of the typical NSD population. For this purpose, we used the 4.5\,$\mu$m extinction map from \citet{Schodel:2014fk}, which provides uniform coverage across the field. At this wavelength, dust extinction is relatively low, allowing us to probe dense clouds and reliably identify high-extinction areas. We analysed the extinction distribution and excluded regions with values above the 70th percentile. In addition, we removed the target areas associated with the 50 and 20\,km/s clouds, as well as the nuclear star cluster, whose stellar population differs from that of the NSD \citep[e.g.][]{Schodel:2020aa,Nogueras-Lara:2021wm,Feldmeier-Krause:2022vm,Chen:2023aa}. The final control region, along with the excluded areas, is shown in the bottom panel of Fig.\,\ref{scheme}.  As a cross-check for our analysis in the following sections, we also used the full region (excluding the shaded contours in the same panel) as a secondary, larger control field.

\subsection{Photometry}
\label{phot}
We used the $HK_s$ photometry from the GALACTICNUCLEUS catalogue \citep{Nogueras-Lara:2019aa}, the state-of-the-art, high-angular-resolution ($\sim0.2''$) $JHK_s$ photometric survey of the NSD, covering its inner $\sim6\,000$\,pc$^2$. The data were acquired with the HAWK-I instrument at the Very Large Telescope \citep{Kissler-Patig:2008fr}, and the catalogue provides point spread function photometry for over 3 million stars, with statistical uncertainties below 0.05\,mag at $H\sim19$ and $K_s\sim18$\,mag. The zero point systematic uncertainties are $\lesssim0.04$\,mag in both bands. The completeness of the catalogue was determined by calculating the critical distance at which a star of any magnitude can be detected around a brighter star \citep{Eisenhauer:1998tg,Harayama:2008ph}. The 80\,\% completeness level is reached at $H\sim18.3$\,mag and $K_s\sim16.3$\,mag \citep[e.g.][]{Nogueras-Lara:2019ad}.

Given the potential saturation of stars with $K_s<11.5$\,mag \citep{Nogueras-Lara:2019aa}, we used the SIRIUS/IRSF survey \citep{Nagayama:2003fk,Nishiyama:2006ai,Nishiyama:2008qa} to replace the photometry of potentially saturated stars, as explained in \citet{Nogueras-Lara:2019ad}. We also included saturated stars that were not originally detected in the GALACTICNUCLEUS catalogue.

\begin{figure}
                 
   \includegraphics[width=\linewidth]{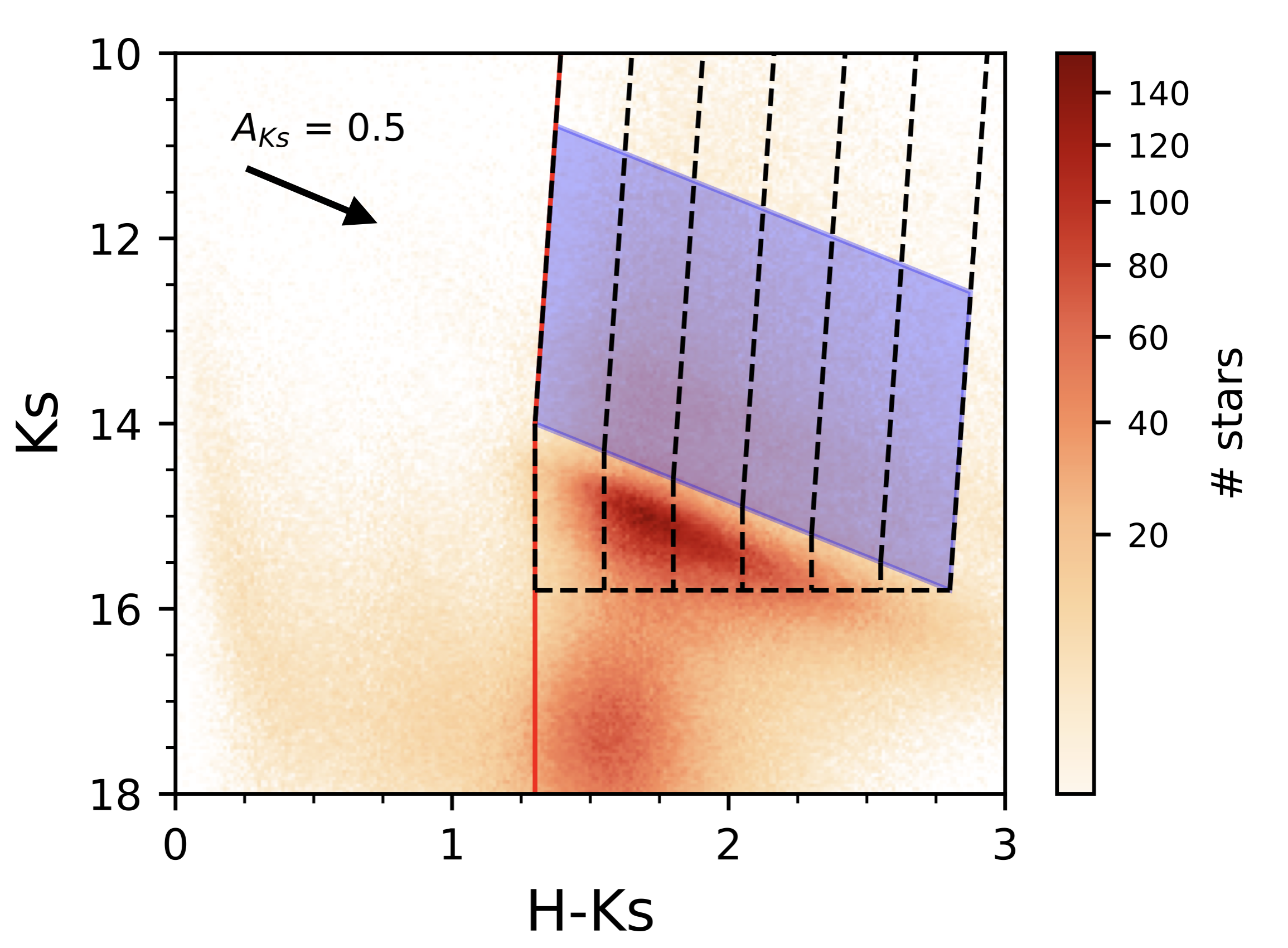}
   \caption{Colour-magnitude diagram $K_s$ versus $H-K_s$ of the stars in the upper panel of Fig.\,\ref{scheme}. The colour scale shows stellar densities using a power stretch scale. The red line indicates the colour cut applied to remove foreground stars (see Sect.\,\ref{phot}). The shaded blue parallelogram indicates the selection box for the proper motion analysis described in Sect.\,\ref{NSD_rot}, whereas the dashed black lines indicate the colour bins for the stellar density analysis in Sect.\,\ref{sect_density}. The black arrow shows the direction of the reddening vector.}

\label{CMD}
\end{figure}

Figure\,\ref{CMD} shows the colour-magnitude diagram $K_s$ versus $H-K_s$ of the region outlined in the upper panel of Fig.\,\ref{scheme}. For all analyses in this paper, we removed the foreground stellar population belonging to the spiral arms along the Galactic centre line of sight and — up to a certain extent — to the Galactic bar/bulge based on significantly different extinction by applying a colour cut  $(H-K_s)>$\,max($1.3, -0.0233K_s+1.63)$ \citep[e.g.][]{Nogueras-Lara:2021wj}, as indicated in Fig.\,\ref{CMD}.

\subsection{Proper motion catalogue}

We conducted our kinematical analysis using the proper motions from the VIRAC2 catalogue \citep{Smith_submitted}. This is an absolute proper motion catalogue based on the multi-epoch Vista Variables in the Via Lactea (VVV) survey \citep{Minniti:2010fk}. The VIRAC2 catalogue uses point spread function photometry and combines the $K_s$ photometry from VVV and VVVX \citep[the temporal extension of the VVV catalogue;][]{Saito:2024aa} to build an absolute proper motion catalogue based on the Gaia reference frame. To obtain proper motions relative to the Galactic centre, we applied a first-order correction by subtracting the proper motion of Sgr\,A$^*$ \citep[$\mu_l=-6.396\pm0.071$\,mas/yr and $\mu_b=-0.239\pm0.045$\,mas/yr in Galactic coordinates;][]{Gordon:2023aa}.

We chose stars with proper motion components parallel and perpendicular to the Galactic plane, $\mu_l$ and $\mu_b$, within $|\mu_l|, |\mu_b| < 15$\,mas/yr to select NSD stars while filtering out those with high proper motions \citep[e.g.][]{2015ApJ...812L..21S,Shahzamanian:2021wu}. We crossmatched the proper motion catalogue with the GALACTICNUCLEUS photometry in the target region (field of view shown in Fig.\,\ref{scheme}), after removing foreground stars as previously described in Sect.\,\ref{phot}, using a search radius of $\sim0.2''$ (i.e. the angular resolution of the GALACTICNUCLEUS survey).

We restricted our analysis to stars with $K_s > 11$\,mag to avoid bright stars that are saturated in the VVV images and show reduced astrometric precision \citep[see Sect.\,3.2 in][]{Smith_submitted}. Additionally, we excluded faint sources by applying a lower magnitude cut parallel to the reddening vector, to avoid biases due to varying extinction across regions at different distances along the line of sight. The final selection is shown as the shaded blue region in Fig.\,\ref{CMD}, and comprises $\sim80\,000$ stars. The mean proper motion uncertainty of the sample is $\sim1$\,mas/yr.

To assess the completeness of the proper motion dataset, we built a $K_s$ luminosity function of the stars with measured proper motions and compared it to a similar luminosity function for all GALACTICNUCLEUS stars in the region, following the process described in \citet{Nogueras-Lara:2023ab}. We determined a completeness level of $\gtrsim 60$\,\% at $K_s\sim16$\,mag for stars with proper motions, based on the GALACTICNUCLEUS survey being approximately fully complete at this magnitude \citep[e.g.][]{Nogueras-Lara:2019ad}.

\begin{figure*}
                 
   \includegraphics[width=\linewidth]{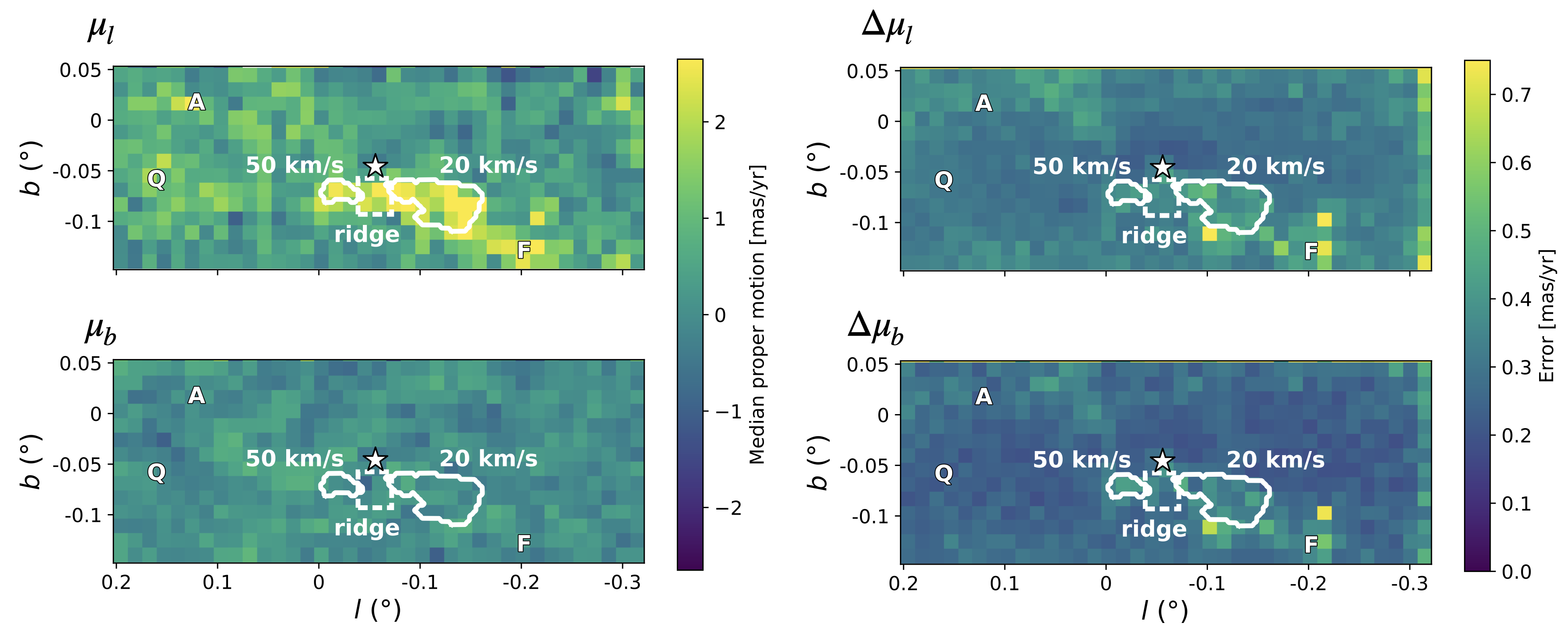}
   \caption{Median proper motion maps (first column) and corresponding uncertainty maps (second column). The positions of the 50 and 20\,km/s clouds are indicated, along with a region of high $\mu_l$ values marked as the `ridge'. The white star marks the position of Sgr\,A*, and `A' and `Q'  denote the Arches and Quintuplet clusters \citep[e.g.][]{Clark:2018ab,Rui:2019aa}. The `F' marks a region showing higher proper motions than its surroundings, probably due to a dark cloud in front of the CMZ. Colour scales are set independently for the median and uncertainty panels to match the range of their respective values.}

\label{mu_map}
\end{figure*}

\section{Stellar kinematics}

The NSD is a rotating structure \citep[e.g.][]{2015ApJ...812L..21S,Shahzamanian:2021wu,Nogueras-Lara:2022aa} that shows differential rotation. Stars on the near and far edges rotate eastwards and westwards, respectively, while the rotation slows towards the central regions \citep{Nogueras-Lara:2023ab}. A statistical correlation exists between extinction and distance along the line of sight \citep{Nogueras-Lara:2022aa,Nogueras-Lara:2023aa}. Stars near the front edge of the NSD typically show lower extinction and faster rotation than those located deeper inside, where the extinction is higher \citep{Martinez-Arranz:2022uf,Nogueras-Lara:2023ab}. This correlation allows $\mu_l$, which traces NSD rotation, to be used to infer the position of a cloud along the line of sight. The presence of a cloud increases extinction for stars within or behind it, obscuring them. This leads to a different proper motion distribution compared to nearby regions with lower extinction, which sample deeper layers of the NSD and thus display a different average rotational motion.

\subsection{Proper motion maps}
\label{prop_map}

We built $\mu_l$ and $\mu_b$ maps using a pixel size of approximately $50''\,\times\,50''$, equivalent to $\sim$2\,pc$\,\times\,$2\,pc at the Galactic centre distance. For each pixel, we calculated the median proper motion of the stars within it. Median values were chosen over mean values to account for potential non-Gaussian distributions in the data. Additionally, we generated uncertainty maps by calculating the standard error of the median for each pixel. Figure\,\ref{mu_map} presents the resulting proper motion maps along with their associated uncertainties.

The $\mu_l$ map reveals higher proper motion values in the regions of the 50 and 20\,km/s clouds compared to their surroundings. This suggests that these areas are dominated by stars rotating eastwards, typical of the near side of the NSD \citep[e.g.][]{Nogueras-Lara:2023ab}. In contrast, the surrounding regions show $\mu_l$ values closer to zero, reflecting a mix of stars located at different depths within the NSD. We interpret this as evidence that the clouds are located on the front side of the NSD along the line of sight, obscuring stars from deeper layers, where proper motions are smaller, close to zero, or even westwards, depending on their position within the NSD. Furthermore, we identified a region of high $\mu_l$ connecting the two clouds (see Fig.\,\ref{mu_map}), which suggests a possible physical link between them. This feature likely corresponds to the `molecular ridge' previously reported as a bridge of material connecting the clouds \citep{Ho:1991aa,Herrnstein:2005aa}. We refer to this region as the `ridge' in the following.

The $\mu_l$ map also shows a region of high proper motions to the south-west of the 20\,km/s cloud. Nevertheless, this region does not coincide with a prominent H$_2$ column density and/or a known CMZ cloud \citep[e.g.][]{Henshaw:2022vl,Battersby:2024aa}. We therefore believe that it traces a foreground cloud along the line of sight towards the Galactic centre that lies close in projection to the 20\,km/s cloud, but is not physically connected to it. The $\Delta\mu_l$ map also shows higher relative uncertainties associated with this region, probably indicating the presence of stars with significantly different $\mu_l$ values, as expected if the cloud is a foreground feature obscuring stars that are not part of the Galactic centre region.

By comparison, the $\mu_b$ map appears homogeneous, showing no significant differences between the 50 and 20\,km/s clouds and the surrounding regions. This result is expected, as the proper motion component perpendicular to the Galactic plane does not trace the NSD rotation and is therefore unaffected by the presence of clouds.

\subsection{Proper motion distribution}
\label{pm_dist}

We also analysed the proper motion distributions of individual stars in the regions associated with the 50 and 20\,km/s clouds, as well as in the ridge, and compared them with those from the control field (see the lower panel in Fig.\,\ref{scheme}). We derived the $\mu_l$ and $\mu_b$ distributions and used the Gaussian mixture model (GMM) function from the scikit-learn Python library \citep[][]{Pedregosa:2011aa} to model the underlying probability density function with a mixture of Gaussians. Following previous work \citep[e.g.][]{Martinez-Arranz:2022uf,Nogueras-Lara:2022aa}, we tested models with up to three Gaussians for the $\mu_l$ distributions and up to two for the $\mu_b$ distributions.

\begin{figure}
                 
   \includegraphics[width=\linewidth]{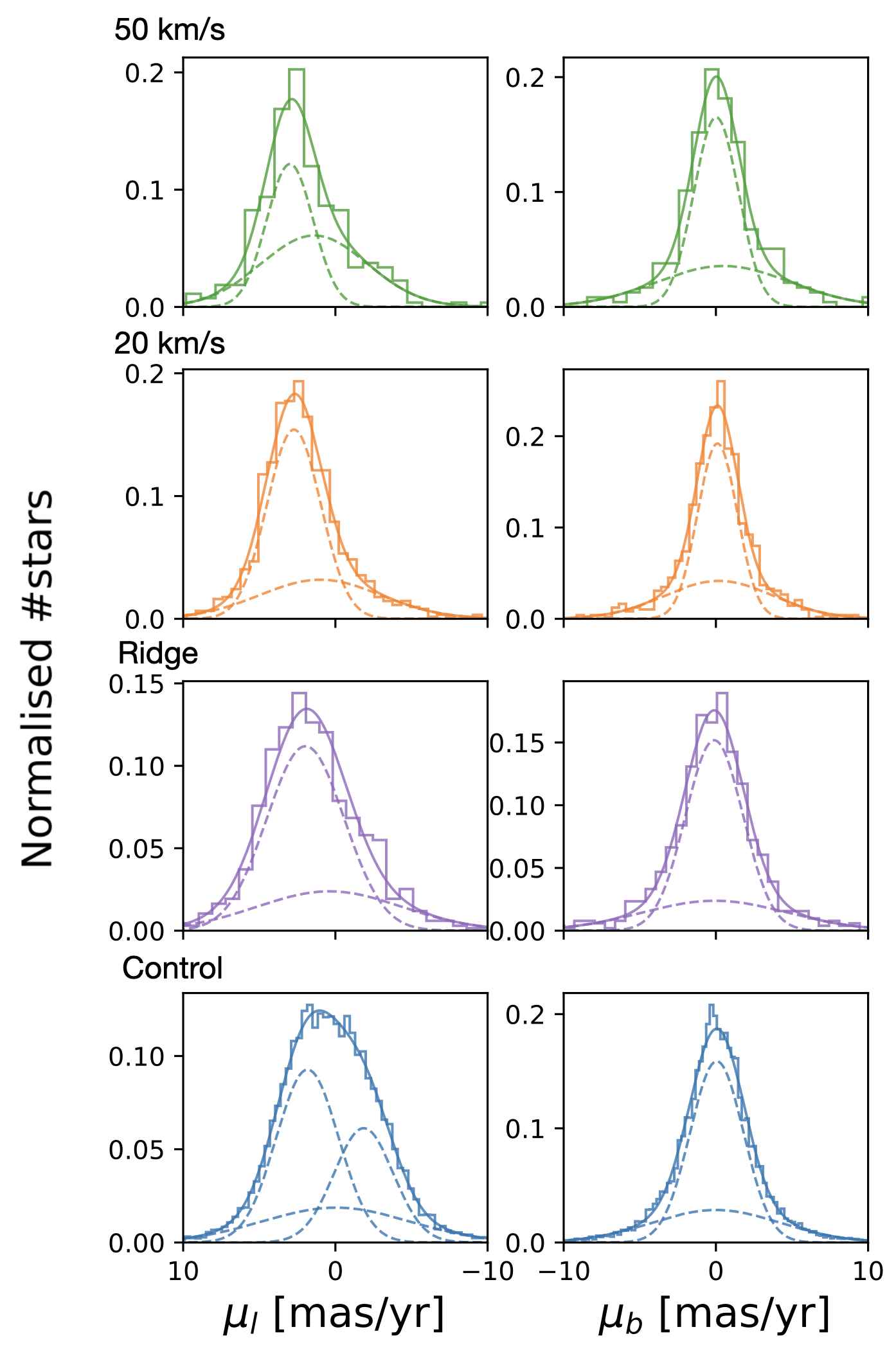}
   \caption{GMM decomposition of $\mu_l$ and $\mu_b$ for stars in the regions associated with the 50 and 20\,km/s clouds, the ridge, and the control field. Dashed lines indicate the individual Gaussian components, and the solid line shows the combined model.}

\label{proper_dist}
\end{figure}

We used the Bayesian information criterion \citep[][]{Schwarz:1978aa} to identify the best model for the data. For the $\mu_l$ distributions, two Gaussians provided the best representation of the 50 and 20\,km/s cloud regions and the ridge, while three were needed for the control field. For the $\mu_b$ distributions, two Gaussians sufficed in all cases. Figure\,\ref{proper_dist} and Table\,\ref{GMM} show the obtained results. To minimise the influence of initial seeds, we performed 1\,000 random initialisations for each GMM within the range [\,-3, 3\,]\,mas/yr \citep[according to previous results for the NSD; e.g.][]{Martinez-Arranz:2022uf,Nogueras-Lara:2023ab} for both the $\mu_l$ and $\mu_b$ components.

To estimate the associated uncertainties, we performed Monte Carlo simulations by generating 1\,000 realisations of the $\mu_l$ and $\mu_b$ distributions for each of the analysed regions. Each synthetic dataset was created by adding Gaussian noise to the measured proper motions, using the individual uncertainties as the standard deviation of the noise. For each realisation, we repeated the GMM calculation, assuming the number of Gaussians determined from the original data. The final uncertainties were computed as the standard deviation of the resulting values (Table\,\ref{GMM}).

Previous analyses of the central region of the NSD showed that the $\mu_l$ distribution is described by three Gaussians: one for eastward moving stars, one for westward moving stars, and an additional Gaussian centred around zero that accounts for the Galactic bar/bulge contamination, which contributes approximately 30\,\% in the central fields \citep[e.g.][]{Martinez-Arranz:2022uf,Sormani:2022wv}. This matches our results for the control field, where the Gaussian representing westward moving stars is smaller than that for eastward moving stars, likely due to the lower completeness of stars from the far side of the NSD, where extinction is higher on average \citep[e.g.][]{Martinez-Arranz:2022uf,Nogueras-Lara:2022aa}. This lower completeness does not affect our conclusions, since our goal here is to compare the number of Gaussian components that best describe the $\mu_l$ distributions of the clouds and the ridge with that of the control field. In these regions, the $\mu_l$ distributions are best modelled with two Gaussian components. The absence of the third Gaussian component — corresponding to stars rotating westwards on the far side of the NSD — suggests that the clouds obscure most of these stars. Only a small fraction of those stars appears to contribute to the central Gaussian, whose relative weight is somewhat higher than in the control field. This indicates that the $\mu_l$ distributions in the three analysed regions are dominated by stars on the near side of the NSD, where the clouds are likely located.

On the other hand, the $\mu_b$ distributions are always best modelled by two Gaussians. The narrow, dominant Gaussian corresponds to NSD stars, while the broader one likely arises from contamination by the Galactic bar/bulge, consistent with the expected contamination level \citep[i.e. $\sim30$\,\%; see Table\,2 in][]{Sormani:2022wv}.

\subsection{NSD rotation and extinction}
\label{NSD_rot}

\begin{figure}
\centering
                 
   \includegraphics[width=\linewidth]{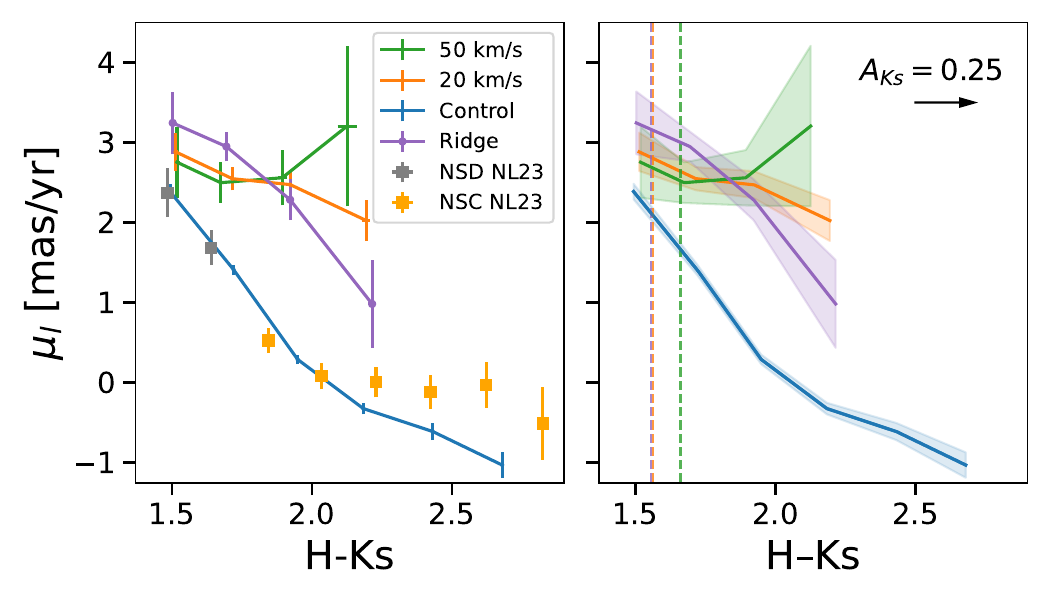}
   \caption{Left panel: $\mu_l$ as a function of $H-K_s$, used as an extinction proxy. Error bars indicate the uncertainties for each data point. Grey and orange points show the results obtained by \citet{Nogueras-Lara:2023ab} along the line of sight towards the nuclear star cluster (NSC; see the legend). Right panel: Interpolated values and associated uncertainties. The vertical lines mark the $H-K_s$ values at which the $\mu_l$ of the cloud regions differ by 3$\sigma$ from that of the control field. The black arrow shows the direction of increasing extinction with $H-K_s$.}

\label{mul_colour}
\end{figure}

To further investigate the 3D position of the 50 and 20\,km/s clouds along the line of sight, as well as their potential physical connection through the ridge, we analysed the dependence of $\mu_l$ on reddening, using the $H - K_s$ colour as a proxy for extinction \citep[e.g.][]{Nogueras-Lara:2023ab}. We divided the proper motion selection box into colour bins of $0.25$\,mag (see Fig.\,\ref{CMD}). For each bin, we computed the median $\mu_l$ values in the regions corresponding to the 50 and 20\,km/s clouds, the ridge, and the control field. We estimated the uncertainties using a Monte Carlo approach that incorporates both sample variance and measurement uncertainties. We performed 10\,000 iterations in which the data were resampled with replacement and each value was perturbed according to its individual uncertainty. The standard deviation of the resulting distribution of medians was taken as the final uncertainty. Similarly, we estimated the median $H-K_s$ values and their associated uncertainties. Figure\,\ref{mul_colour} and Table\,\ref{table2} show the obtained results. We did not include values beyond $H-K_s\sim2.2$\,mag for the clouds and the ridge regions due to the low number of stars and the high associated uncertainties that make the values unreliable.

The results for the control field trace the rotation of the NSD and confirm its dependence on extinction. Less reddened stars exhibit $\mu_l>0$\,mas/yr (i.e. eastward rotation), which decreases with increasing extinction. The rotation curve is not centred at zero because the lower completeness of westward rotating stars (which are fainter due to higher extinction) affects the median $\mu_l$ values. Additionally, the NSD extends beyond a colour of $H-K_s \sim 2.7$\,mag. To assess our results, we overplotted those obtained by \citet{Nogueras-Lara:2023ab} along the line of sight towards the nuclear star cluster, which are based on the proper motions from \citet{Shahzamanian:2021wu}. These include two points corresponding to the NSD and others belonging to the nuclear star cluster. We find excellent agreement between our control-field measurements and the NSD points from \citet{Nogueras-Lara:2023ab}, while the nuclear star cluster points exhibit a distinct rotation, particularly evident at redder colours.

The $\mu_l$ values of stars in the 50 and 20\,km/s cloud regions at the edge of the NSD ($H - K_s \sim 1.5$\,mag) are consistent, within uncertainties, with those of the control field. In contrast, the ridge exhibits a slightly higher $\mu_l$ value, though still compatible with that of the clouds region.

At a similar colour of $H - K_s \gtrsim 1.6$\,mag, the $\mu_l$ values of the clouds and the ridge differ from that of the control field, suggesting that stars in the control field rotate more slowly than those in the cloud and the ridge regions. At redder colours, the median $\mu_l$ values in the clouds and in the ridge remain systematically higher than in the control field.

We interpret this as the clouds obscuring stars from larger NSD radii, making them appear redder than the average control stars at the same depth. The observed kinematic differences indicate that the clouds start to influence the stellar population from $H - K_s \sim 1.6$\,mag onwards, placing them in the nearer half of the NSD. Their position in colour space also suggests that they lie in front of the nuclear star cluster \citep[e.g.][]{Launhardt:2002nx,Schodel:2014bn,Feldmeier:2014kx}, which begins to appear at $H - K_s \sim 1.8$\,mag  (see the nuclear star cluster points from \citet{Nogueras-Lara:2021wm} in Fig.\,\ref{mul_colour}. The significantly higher $\mu_l$ values of the clouds compared to those of the nuclear star cluster stars further support this interpretation. Moreover, the similar trends observed in the 50 and 20\,km/s cloud regions and the ridge suggest they are located at comparable depths within the NSD, supporting a physical connection between the two clouds through the ridge.  The slightly lower $\mu_l$ values measured for the ridge at $H-K_s\sim2.25$\,mag, compared to those of the 50 and 20\,km/s clouds, are consistent with the ridge containing less dense molecular gas, as shown in Fig.\,2 of \citet{Battersby:2024aa}. This would result in lower extinction, allowing more stars located behind the ridge — characterised by the lower $\mu_l$ values typical of stars deeper inside the NSD — to contribute to the median $\mu_l$ at redder $H-K_s$ colours.

To identify the $H-K_s$ values of the clouds and the ridge at which the $\mu_l$ value differs significantly (3$\sigma$) from that of the control field, we interpolated their $\mu_l$ values and corresponding uncertainties, as shown in the right panel of Fig.\,\ref{mul_colour}. Table\,\ref{3sigma} lists the obtained $H-K_s$ values together with the corresponding $\mu_l$ values for the control region. The associated uncertainties were estimated by varying the colour bin widths (0.2 and 0.3\,mag) and by using a larger control field that includes the entire low-extinction region shown in Fig.\,\ref{scheme}, as described in Sect.\,\ref{data}.

\setlength{\tabcolsep}{5.5pt}
\def\arraystretch{1.2}

\begin{table}[h]

\caption{$H-K_s$ values at which the $\mu_l$ values of the cloud regions differ by 3$\sigma$ from the control region.}

\label{3sigma} 
\begin{center}

\begin{tabular}{ccc}
 &  & \tabularnewline
\hline 
\hline 
 & $H-K_s$ & $\mu_{l\ control}$\tabularnewline
 & mag & mas/yr\tabularnewline
\hline 
50 km/s & 1.66$\pm$0.04 & 1.67$\pm$0.08\tabularnewline
20 km/s & 1.56$\pm$0.04 & 2.09$\pm$0.10\tabularnewline
Ridge & 1.56$\pm$0.03 & 2.12$\pm$0.09\tabularnewline
\hline 
 &  & \tabularnewline
\end{tabular}

\end{center}
\footnotesize
\textbf{Notes.} $\mu_{l\ control}$ corresponds to the median $\mu_l$ values of the control field at which the cloud regions differ by 3$\sigma$.
\end{table}

\section{Stellar density}
\label{sect_density}

We used the GALACTICNUCLEUS $HK_s$ photometry  (Fig.\,\ref{scheme}) to investigate the stellar density distribution across the 50 and 20\,km/s molecular clouds and their surroundings. We analysed stellar densities as a function of $H-K_s$ colour (i.e. extinction) to identify the position of the clouds in colour space, based on a decrease in relative stellar density.

We defined $H-K_s$ bins of $0.25$\,mag (see Fig.\,\ref{CMD}). To ensure a completeness $\gtrsim80$\,\% for all colour bins, we set a threshold of $K_s=15.8$\,mag and restricted the analysis to five colour bins. Given the larger number of available stars, compared to the proper motion map in Sect.\,\ref{prop_map}, we created a finer map with a pixel size of $\sim25''\times25''$, corresponding to $\sim1$\,pc$^2$ at the Galactic centre distance.

\begin{figure}

   \includegraphics[width=\linewidth]{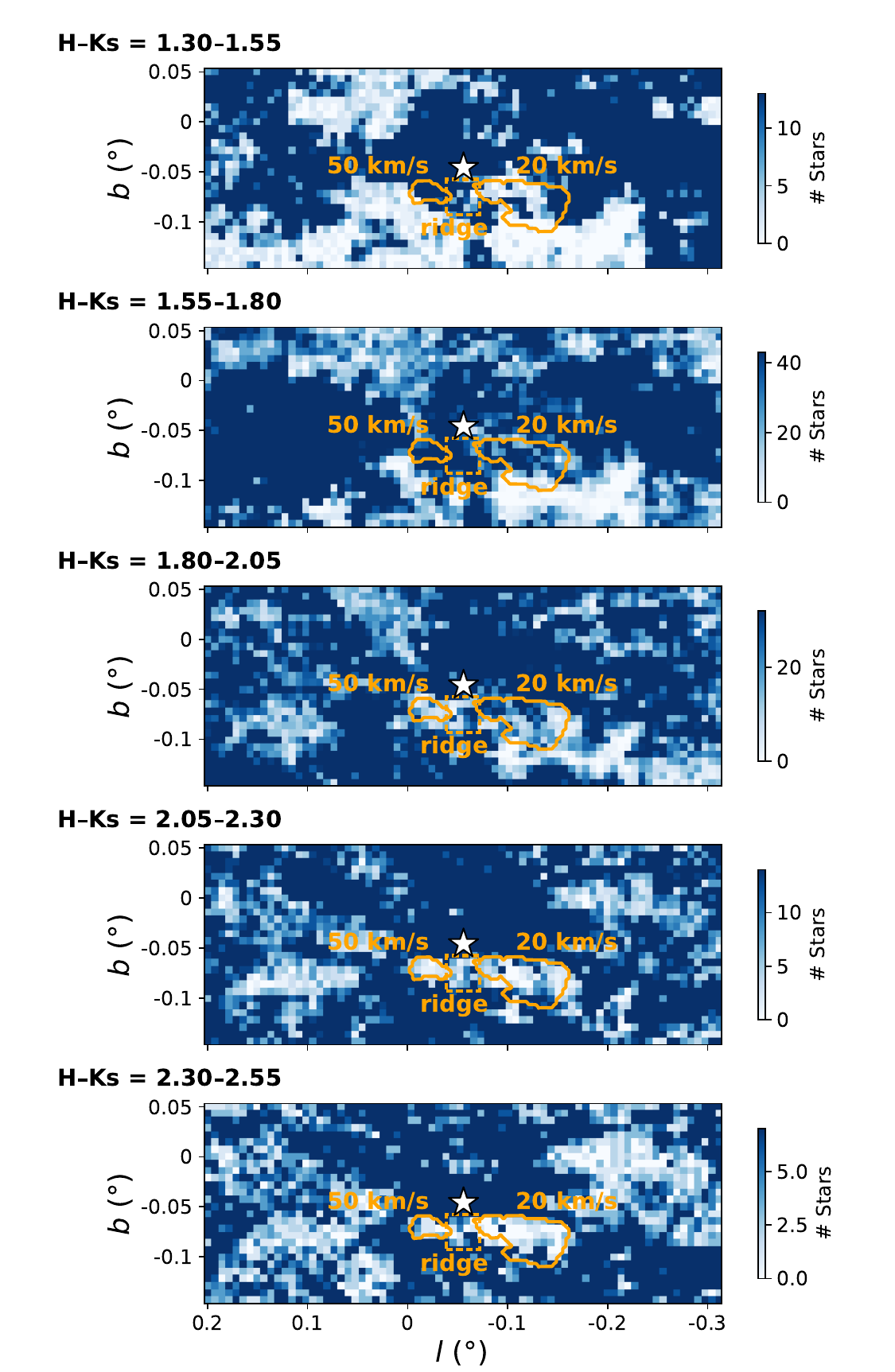}
   \caption{Stellar density maps of the target region for different colour cuts, with increasing extinction from top to bottom. Each panel indicates the  range of the corresponding colour cut. The 50 and 20\,km/s clouds are marked with orange circular regions, and the ridge is indicated by a dashed orange rectangle. The white star in each panel indicates the position of Sgr\,A*. The colour scale represents the number of stars per pixel and is adapted in each panel to the number of stars in the 50 and 20\,km/s clouds for an easier comparison with the surrounding area.}

\label{density}

\end{figure}

Figure\,\ref{density} shows the resulting maps, where the $H-K_s$ colour range of the stars in each bin is indicated. The maps reveal a patchy structure consistent with extinction variations in the NSD on arcsecond scales \citep[e.g.][]{Launhardt:2002nx,Battersby:2020vt,Nogueras-Lara:2021wj}. The first density map ($H-K_s\in[1.3,1.55]$\,mag) displays several low-density regions, which we interpret as foreground clouds along the line of sight, located in front of the NSD. This is consistent with the extinction maps published by \citet[see their Fig.\,3]{Nogueras-Lara:2021wj}, where the first extinction layer maps contain empty pixels in approximately the same locations due to the lack of stars available to estimate an extinction value. Part of this foreground extinction pattern particularly affects the regions of the 20\,km/s cloud and the ridge, where the stellar density is lower than in the surrounding area. This region partially corresponds to the foreground cloud `F' outlined in Fig.\,\ref{mu_map} and discussed in Sect.\,\ref{prop_map}.

Dense clouds, such as the 50 and 20\,km/s ones, increase extinction, reddening background stars and, if sufficiently opaque, blocking their light entirely. A visual inspection of Fig.\,\ref{density} shows that the relative stellar density decreases at $H-K_s \in [1.80,2.05]$\,mag. At redder colours, both clouds appear as regions with very few stars, indicating that they are obscuring background stars. Additionally, the ridge also appears as a low density region connecting the clouds.

This suggests that both clouds and the ridge span a similar range in colour/extinction space, consistent with comparable NSD radii. Their presence in front of stars with $H-K_s \in [1.80,2.05]$\,mag implies that they already obscure background sources at these colours. This threshold is slightly redder than the one inferred from the $\mu_l$ distribution (Sect.\,\ref{NSD_rot}), but fully consistent with it. Compared to the proper motion analysis, the stellar density method traces the effect of the clouds only once they are dense enough to fully block background light and cause a measurable drop in source counts. On the other hand, the proper motion distribution begins to shift at bluer colours because the first stars affected by extinction — typically NSD stars with higher $\mu_l$ — are displaced to redder bins. This changes the motion profile and reveals the presence of the clouds before their impact becomes evident in the stellar densities.

\section{Line-of-sight distance to the clouds}
\label{dist_est}

The correlation between mean extinction and distance along the line of sight does not enable precise distance estimates, as extinction can vary non-uniformly along the line of sight \citep[e.g.][]{Nogueras-Lara:2021wj}. However, assuming that the NSD is axisymmetric, we can use a dynamical model of the NSD to estimate the line-of-sight distance of the clouds given the $\mu_l$ measured in Sect.\,\ref{NSD_rot}.

\begin{figure}

   \includegraphics[width=\linewidth]{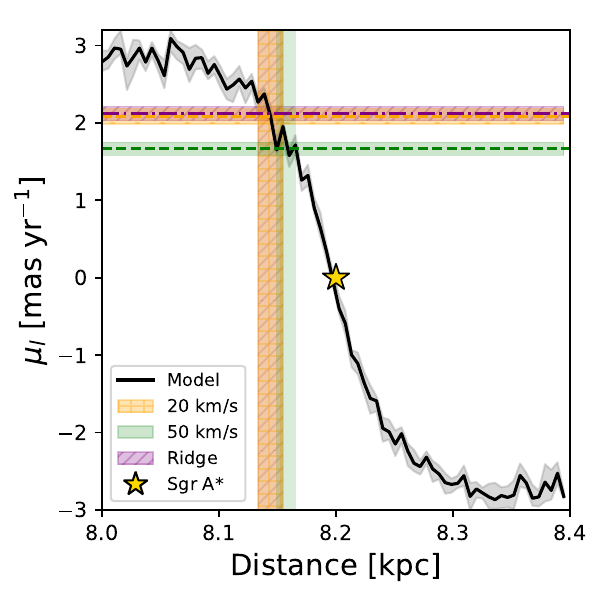}
   \caption{Theoretical $\mu_l$ profile obtained for the control region in Fig.,\ref{scheme}. The grey shaded region indicates the model uncertainty. The position of Sgr\,A$^*$ (8.2\,kpc and 0\,mas/yr) is marked by the yellow star. The orange, green, and purple horizontal lines indicate the $\mu_l$ values at which the clouds are detected ($\mu_l$ differing from that of the control field at the 3$\sigma$ level), with their uncertainties shown as shaded regions. The corresponding estimated distance ranges of the clouds are shown as vertical shaded regions.}

\label{profile}
\end{figure}

We used an N-body realisation of the axisymmetric equilibrium model of the NSD from \citet{Sormani:2022wv}, generated with AGAMA \citep{Vasiliev:2019aa}. We created the velocity profile $\mu_l$ versus distance by selecting simulated stars in a region equivalent to the control field (red box in Fig.\,\ref{scheme}). We then applied bootstrap resampling with replacement in each distance bin to estimate the corresponding uncertainties, defined as the 16th and 84th percentiles. Figure\,\ref{profile} shows the resulting theoretical $\mu_l$ profile.

To estimate the distance of the clouds and ridge regions along the line of sight, we projected the $\mu_l$ values of the control field obtained in Sect.\,\ref{NSD_rot} (see Table\,\ref{3sigma}) onto the theoretical $\mu_l$ profile. The vertical shaded regions in Fig.\,\ref{profile} show the range of distances at which the presence of the clouds is detected kinematically. We find that the 50 and 20\,km/s clouds and the ridge are located at $43\pm8$\,pc, $56\pm11$\,pc, and $56\pm11$\,pc from Sgr\,A$^*$, respectively. The estimated distance uncertainties account for both the observational $\mu_l$ uncertainties and those of the theoretical model.

\begin{figure}
              
   \includegraphics[width=\linewidth]{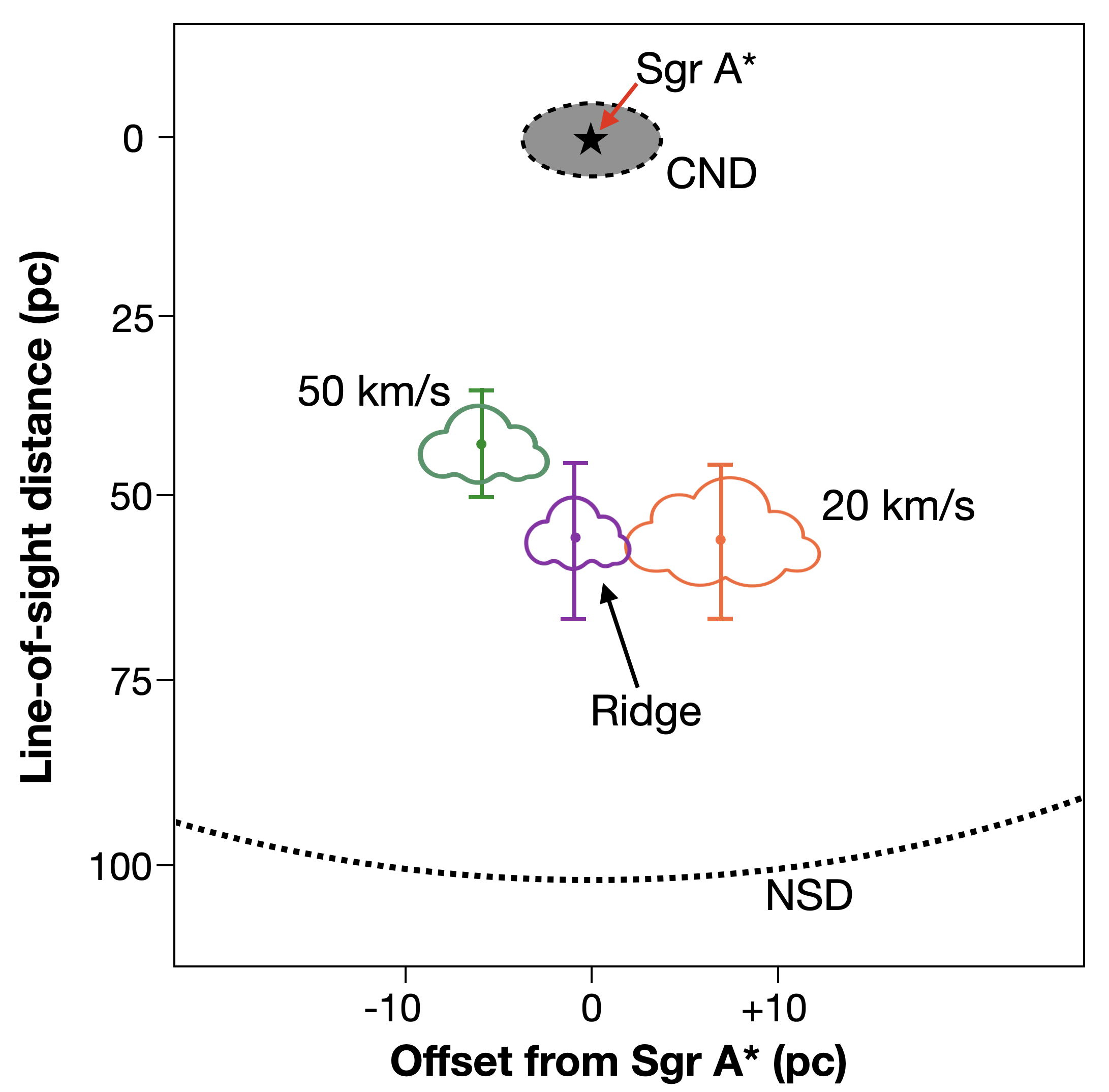}
   \caption{Top-down schematic view of the line-of-sight positions of the 50 and 20\,km/s clouds and the ridge,  inferred from the self-consistent NSD model of \citet{Sormani:2022wv} and the $\mu_l$ versus $H-K_s$ relation explained in Sect.\,\ref{NSD_rot}. This plot assumes that the position of the clouds and the ridge is that at which their $\mu_l$ values are 3\,$\sigma$ different from that of the control field. The x- and y-axis scales differ to optimise the visualisation of the relative distances. The positions of Sgr\,A$^*$ and the circumnuclear disc (CND) are also shown. The dotted black line marks the scale length of the NSD \citep[e.g.][]{gallego-cano2019,Sormani:2022wv}.
}

\label{scheme_distance}
\end{figure}

Figure\,\ref{scheme_distance} displays a schematic top view of the inferred line-of-sight positions. The derived distances for the clouds and the ridge are consistent within the uncertainties, suggesting a probable physical connection among them.

However, this method should be interpreted with caution, as it is based on a rough approximation that assumes an axisymmetric NSD structure. In addition, our estimate uses the $H-K_s$ colour at which the $\mu_l$ values differ significantly (3$\sigma$) from those of the control field, so it should be regarded as a lower limit on the clouds’ distance from Sgr\,A*. Hence, we cannot exclude that the clouds lie somewhat further out within the near side of the NSD, with the similar $\mu_l$ values for the clouds and the control field at $H-K_s\sim1.5$\,mag (particularly when adopting the values from \citealt{Nogueras-Lara:2023ab}), suggesting that they are not detected too close to the outer edge of the NSD in the proper motion analysis.

The $\mu_l$ versus $H-K_s$ relation used for the control field (Fig.\,\ref{mul_colour}) is based on median values, which may oversimplify a more complex underlying distribution (see Fig.\,\ref{proper_dist}). In addition, the model does not include any completeness correction, although the control field is expected to be highly complete, at least for the front side of the NSD. Finally, the line-of-sight extent of the clouds is unknown, precluding an estimate of their spatial depth; as a result, we can only provide an approximate location, particularly where they begin to obscure background stars.

\section{Stellar populations}

There are indications of an age gradient in the NSD, suggesting an inside-out formation scenario. This is consistent with gas being channelled to the Galactic centre by the Galactic bar \citep[e.g.][]{Baba:2020aa}, as proposed for nearby Milky Way-like galaxies \citep[e.g.][]{Bittner:2020aa}. The innermost regions of the NSD are dominated by an old stellar population, with over 80\,\% of the stellar mass formed more than 7\,Gyr ago \citep{Nogueras-Lara:2019ad,Sanders:2022ab,Nogueras-Lara:2022ua,Schodel:2023aa}. In contrast, a significant fraction of intermediate-age stars are found near the edge of the NSD, where $\sim40$\,\% of the stellar mass has ages between 2 and 7\,Gyr \citep{Nogueras-Lara:2023ab,Nogueras-Lara:2024aa}. Analysing the stellar populations towards the 50 and 20\,km/s clouds, and the ridge can therefore provide insights into their position along the line of sight.

\subsection{De-reddened $K_s$ luminosity functions}

To study the stellar populations towards the target regions, we constructed de-reddened $K_s$ luminosity functions (using the GALACTICNUCLEUS photometry after removing the foreground population, as explained in Sect.\,\ref{phot}), and fitted them with theoretical models, following the method described in \citet{Nogueras-Lara:2022ua}. We first built an extinction map using red clump stars \citep[core-helium-burning stars with well-characterised intrinsic properties;][]{Girardi:2016fk} and red giants with similar colours, and used it to correct the photometry for extinction. We then derived de-reddened $K_s$ luminosity functions for the regions corresponding to each cloud and the control field by selecting stars with $H - K_s \lesssim 1.8$\,mag. This colour threshold ensures that we probe stars at the line-of-sight distances where the clouds are expected to lie, based on our previous analysis (Sects.\,\ref{NSD_rot} and \ref{sect_density}).

We also accounted for the impact of crowding on completeness, as it is the primary cause of incompleteness at the Galactic centre due to the extreme stellar densities \citep[e.g.][]{Nogueras-Lara:2019ad,Schodel:2023aa}. Using the method outlined by \citet{Eisenhauer:1998tg} and \citet{Harayama:2008ph}, we computed dedicated completeness solutions for each of the target regions, and assumed a completeness uncertainty of $\sim5$\,\% for a completeness level of $\gtrsim90$\,\%.

We applied these completeness corrections to the respective $K_s$ luminosity functions, imposing a lower limit of $\sim90\,\%$ completeness. This threshold corresponds to the point where the $K_s$ luminosity functions become sensitivity-limited, indicated by a sharp decrease in the number of stars beyond this limit.

\subsection{Fit of the $K_s$ luminosity functions}

We analysed the $K_s$ luminosity functions by fitting them with a linear combination of theoretical stellar population models, using the approach in \citet{Nogueras-Lara:2019ad,Nogueras-Lara:2022ua}, and \citet{Schodel:2023aa}. The fitting included a Gaussian smoothing parameter to account for potential variation in the distances of stars \citep[the NSD scale length is $\sim100$\,pc,][]{gallego-cano2019,Sormani:2022wv}, residual reddening, and also a parameter to account for the distance to the Galactic centre.

We employed PARSEC\footnote{generated by CMD 3.6 (\url{http://stev.oapd.inaf.it/cmd})} \citep{Bressan:2012aa,Chen:2014aa,Chen:2015aa,Tang:2014aa,Marigo:2017aa,Pastorelli:2019aa,Pastorelli:2020wz} and MIST models \citep{Paxton:2013aa,Dotter:2016aa,Choi:2016aa} with twice solar metallicity \citep[e.g.][]{Schultheis:2019aa,Schultheis:2021wf,Feldmeier-Krause:2022vm}, and similar stellar ages (14, 11, 8, 6, 3, 1.5, 0.6, 0.4, 0.2, 0.1, 0.04, 0.02, 0.01, and 0.005\,Gyr). The ages were chosen to capture significant variations in the shape of the $K_s$ luminosity function that are more relevant for younger populations \citep[see the supplementary material in][]{Nogueras-Lara:2022ua}.

\begin{figure*}
                 
   \includegraphics[width=\linewidth]{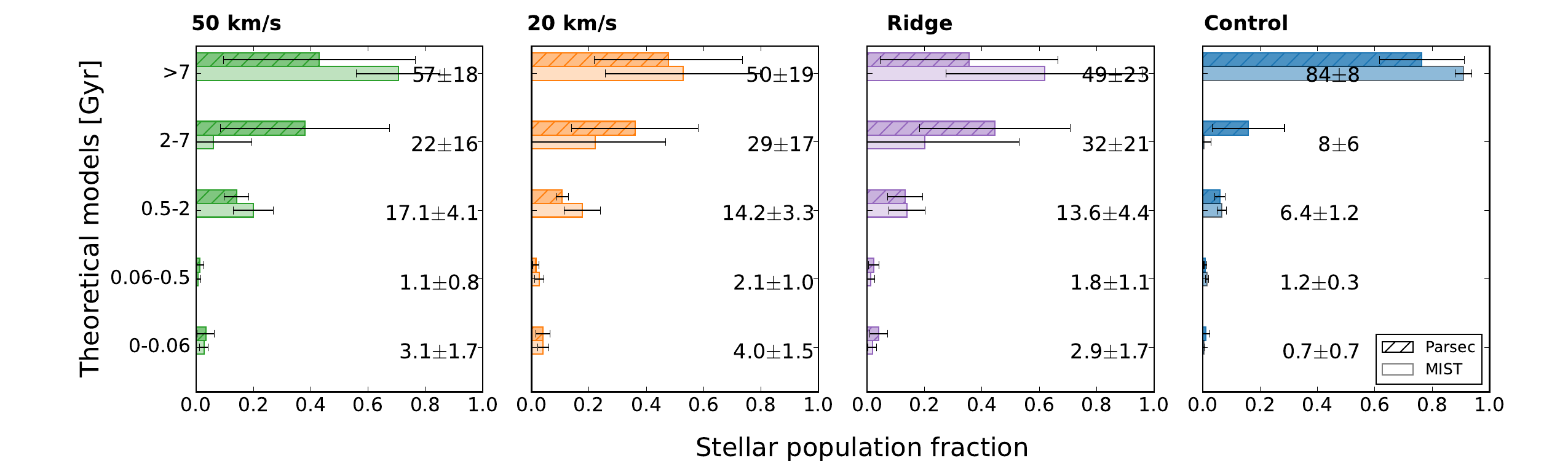}
   \caption{Results of the $K_s$ luminosity function fitting for the 50 and 20\,km/s clouds, the ridge, and the control field regions. The numbers in each panel represent the percentage contribution of each age bin, with the associated uncertainty calculated as the average of the results from the PARSEC and MIST models (as indicated in the legend).}

\label{res_fit}
\end{figure*}

To estimate the contribution of each stellar model and the associated uncertainty, we generated 1\,000 Monte Carlo simulations of synthetic luminosity functions for each region (the 50 and 20\,km/s clouds, the ridge, and the control field), assuming Gaussian uncertainties. We restricted the bright end of the $K_s$ luminosity functions to $K_s = 8.5$\,mag to mitigate potential saturation effects \citep[e.g.][]{Nagayama:2003fk,Nishiyama:2006ai,Nishiyama:2008qa}. To avoid degeneracies between stellar populations with similar ages, we defined five age bins ($>7$, 2-7, 0.5-2, 0.06-0.5, and 0-0.06\,Gyr), as outlined by \citet{Nogueras-Lara:2022ua}. We then fitted PARSEC and MIST models, calculating the mean contribution of each age bin and determining the uncertainty as the standard deviation. The final results, presented in Fig.\,\ref{res_fit}, were obtained by combining the results from PARSEC and MIST models and quadratically propagating the uncertainties.

Our results reveal the presence of an intermediate-age (2–7\,Gyr) stellar population towards the 50 and 20\,km/s clouds and the ridge. This population is much lower in the control field, which is dominated by an old stellar population, with $\gtrsim80$\,\% of the stellar mass older than 7\,Gyr, consistent with previous findings in the inner region of the NSD \citep[e.g.][]{Nogueras-Lara:2019ad,Schodel:2023aa}. The results towards the 50 and 20\,km/s clouds and the ridge exhibit larger uncertainties compared to the control field, primarily due to the significantly lower number of stars available to construct the luminosity function.

Given that the three $K_s$ luminosity functions span a similar colour range, we would expect a comparable line-of-sight depth across all regions \citep[e.g.][]{Nogueras-Lara:2022aa}, and thus similar stellar populations. However, we obtained the presence of an intermediate-age stellar population towards the 50 and 20\,km/s clouds and the ridge that is barely present in the control field. We interpret this as evidence of the clouds' influence. They redden stars from deeper, inner regions of the NSD, hiding them and thereby enhancing the contribution of the intermediate-age population typically found near the NSD edge. This population becomes more prominent in the studied colour range, where older stars dominate in the case of the control field. As a result, the stellar population towards the 50 and 20\,km/s clouds resembles that observed near the closest edge of the NSD \citep[e.g.][]{Nogueras-Lara:2022ua,Nogueras-Lara:2023aa,Nogueras-Lara:2024aa}. The similar stellar population obtained for the ridge suggests that it also hides stars from the central region of the NSD, and is therefore likely located close to the 50 and 20\,km/s clouds along the line of sight.

Our analysis of the stellar population also allows us to provide a rough estimate of the distance by comparing the stellar populations in the cloud regions and the ridge with those of the NSD at different Galactocentric radii \citep{Nogueras-Lara:2022ua, Nogueras-Lara:2024aa}. In this way, the analysis of the Sgr\,B1 and Sgr\,C fields, located at $\sim80$\,pc from Sgr\,A$^*$, reveals a stellar population similar to that in the cloud regions  and the ridge, suggesting that they might lie at a comparable distance along the line of sight. This is fully consistent with the kinematic estimate presented in Sect.\,\ref{dist_est}.

\subsection{Mass estimate towards the clouds and the ridge}

We estimated the total stellar mass in each region from the fits to the 1\,000 Monte Carlo realisations obtained with the PARSEC models. The resulting masses were normalised by the total area of each region to derive the stellar surface mass density. We obtained total surface masses of $(9.8\pm2.4)\times10^5$\,M$_\odot$ pc$^{-2}$, $(8.2\pm1.5)\times10^5$\,M$_\odot$ pc$^{-2}$, $(6.9\pm1.9)\times10^5$\,M$_\odot$ pc$^{-2}$, and $(2.8\pm0.4)\times10^6$\,M$_\odot$ pc$^{-2}$ for the 50\,km/s and 20\,km/s clouds, the ridge, and the control region, respectively.

For our analysis, we applied the same colour selection to all regions to ensure a consistent comparison of their stellar populations. We find that the control field contains roughly three times more stellar mass per pc$^2$ than the cloud and ridge regions, which are consistent with each other. The lower apparent stellar mass towards the clouds and the ridge can be explained by extinction from dense molecular material obscuring the stars within and behind these regions. The higher stellar mass in the control field supports our interpretation that the clouds are located in the near side of the NSD.

\subsection{Systematic uncertainties}

To evaluate the robustness of our results towards the 50 and 20\,km/s clouds, and the ridge, we conducted a series of tests by varying key parameters that might influence the results:

1. Initial mass function. Our sample consists primarily of giant stars, making our results largely insensitive to variations in the initial mass function \citep{Schodel:2023aa}. For the PARSEC models, we adopted a Kroupa initial mass function \citep{Kroupa:2013wn}, as it accounts for unresolved binaries, while the MIST models used a Salpeter initial mass function \citep{Salpeter:1955qo}. To assess any potential influence of the chosen mass function, we repeated the analysis for the PARSEC models using the Salpeter function and found no significant differences compared to the Kroupa one.

2. Bright and faint end of the $K_s$ luminosity functions. We tested the effect of varying the bright cut to $K_s=8$\,mag, as well as adjusting the faint end completeness limits (87\,\% and 92\,\%, corresponding to $\sim0.5$\,mag variations). The results remained consistent within the uncertainties.

3. Bin width. We experimented with different bin widths for the $K_s$ luminosity functions ($\pm0.05$\,mag). This change did not result in any significant variation in the final outcomes.

4. Metallicity of the stellar population. We repeated the analysis using Parsec models with solar metallicity. This resulted in a slightly smaller contribution from the oldest stellar bin, while the intermediate-age stars (2-7\,Gyr) showed a larger contribution, further supporting our interpretation of the clouds' positions.

5. Extinction map. For our analysis, we constructed an extinction map using reference stars in the same colour range as the $K_s$ luminosity function ($H-K_s \in [1.3,1.8]$\,mag). We also repeated the analysis with an alternative map built from reference stars with $H-K_s \in [1.3,3.0]$\,mag. We found no significant differences, confirming the robustness of our results within the uncertainties.

6. Control field. We repeated the analysis using the entire control region shown in the lower panel of Fig.\,\ref{scheme}. This yields a slightly larger contribution from the oldest stellar population, but no significant difference is found within the uncertainties.

\section{Discussion}

We estimated the line-of-sight positions of the 50 and 20\,km/s clouds in the CMZ and concluded that they are located on the near side of the NSD, with their influence detectable from our kinematic analysis at distances of $43\pm8$\,pc and $56\pm11$\,pc from Sgr\,A$^*$, respectively. The ridge region connecting them in projection is estimated to lie at $56\pm11$\,pc, suggesting that the two clouds are physically connected through the ridge. This conclusion is supported by three independent methods: kinematic analysis, stellar density mapping, and stellar population comparison.

Our findings are in agreement with \citet{Walker:2025aa} and \citet{Lipman:2025aa}, who likewise place the clouds on the near side of the CMZ. Additionally, X-ray observations of CMZ clouds have been used to infer the line-of-sight positions of molecular structures in the Galactic centre \citep[e.g.][]{Capelli:2012aa,Clavel:2013aa,2015MNRAS.453..172P,Stel:2025aa}. Although there is no clear result for the line-of-sight distances of the 50 and 20\,km/s clouds, their lack of strong X-ray illumination suggests that they are located closer to Sgr\,A$^*$ than other CMZ clouds, consistent with our results. Our inferred positions lie between the CMZ model of \citet{Molinari:2011fk}, which situates the clouds at $\lesssim$ 20\,pc from Sgr\,A$^*$, and that of \citet{Kruijssen:2015aa}, which places them at $\gtrsim$ 60\,pc, in agreement with the kinematic analysis of \citet{Henshaw:2016aa}. Nevertheless, because our distance estimate is based on the $H-K_s$ colour at which the $\mu_l$ values show a significant (3\,$\sigma$) deviation from those of the control field, it represents a lower limit on the distance from Sgr\,A*. We therefore cannot exclude that the clouds lie farther from Sgr\,A*, which would still be compatible with the estimate from \citet{Kruijssen:2015aa}. In any case, our conclusion that the clouds are located on the near side of the NSD remains robust.

However, other studies suggest that the clouds lie much closer to the Sgr\,A$^*$ ($\lesssim20$\,pc). This conclusion is based on evidence of interactions with the Sgr\,A East supernova remnant \citep{Genzel:1990aa, McGary:2001aa, Herrnstein:2005aa, Lee:2008aa, Sjouwerman:2008aa}, and the circumnuclear disc \citep{Karlsson:2015aa, Takekawa:2017aa, Tsuboi:2018aa, Ballone:2019aa} — a dense molecular structure located at 1.5–4\,pc from Sgr\,A$^*$ \citep[e.g.][]{Liu:2012aa, Mills:2013aa, Hsieh:2017aa, Hsieh:2021aa}.

While our results constrain the approximate location of the clouds, we cannot determine their full extent along the line of sight. Their projected lengths are $\sim$5\,pc and $\sim$12\,pc for the 50 and 20\,km/s clouds, respectively, so a comparable extent along the line of sight is plausible. However, we cannot rule out the possibility that they are part of elongated gaseous structures — referred to as `feathers' — that extend inwards towards the Galactic centre. This scenario was proposed by \citet{Tress:2020aa} to reconcile models placing the clouds both at the near edge of the NSD and closer to Sgr\,A$^*$. It is also consistent with observations of similar features in the nuclear rings of external galaxies \citep{Benedict:2002aa}. Such a scenario would imply that the 3D geometry of the CMZ is more complex than previously assumed, as suggested by \citet{Tress:2020aa}.

The potential physical connection between the 50 and 20\,km/s clouds has been a subject of debate. Our results for the region connecting them suggest a physical link, consistent with previous studies reporting a molecular ridge between the clouds \citep[e.g.][]{Ho:1991aa, Herrnstein:2005aa}. In contrast, other works interpret them as unrelated structures with no direct association \citep[e.g.][]{Vollmer:2003aa}.

\section{Conclusion}

We investigated the line-of-sight location of the 50 and 20\,km/s clouds in the CMZ using stellar kinematics and photometric data. We built proper motion maps and applied a GMM approach to analyse the $\mu_l$ distribution, finding a consistent excess of eastward moving stars — and a lack of westward moving ones — in the cloud regions and in the ridge connecting them. This indicates that the clouds obscure stars on the far side of the NSD, which rotates westwards, and suggests a potential physical connection between them.

We analysed the dependence of $\mu_l$ on the $H-K_s$ colour, used as a proxy for extinction, and found that the kinematic imprint of the clouds and the ridge emerges at $H-K_s \sim 1.6$\,mag. This was compatible with our analysis of stellar densities at different $H-K_s$ values. The similar behaviour observed for both the clouds and the ridge further supports the idea that they form physically connected structures.

To estimate the position of the clouds, we employed the self-consistent model of the NSD from \citet{Sormani:2022wv} and found that the 50 and 20\,km/s clouds are kinematically detectable (i.e. 3\,$\sigma$ $\mu_l$ difference with respect to the control field) at distances of $43\pm8$\,pc and $56\pm11$\,pc from Sgr\,A$^*$. Similarly, we estimated a distance of $56\pm11$\,pc for the ridge, which is consistent with the two clouds being physically connected through it.

Finally, we analysed the stellar populations towards the cloud regions and the ridge, finding them to be consistent with that of the front side of the NSD \citep{Nogueras-Lara:2022ua, Nogueras-Lara:2024aa}. This also agrees with the interpretation that the clouds are located on the near side and obscure older stars situated deeper within the NSD \citep{Nogueras-Lara:2019ad,Sanders:2022ab}.

Our results are consistent with previous work that placed the clouds on the near side of the NSD \citep{Kruijssen:2015aa, Henshaw:2016aa}. Nevertheless, we cannot rule out the possibility that the clouds are elongated gaseous structures extending towards the central few parsecs, as proposed by \citet{Tress:2020aa}.

\begin{acknowledgements}

This work is based on observations made with ESO Telescopes at the La Silla Paranal Observatory under program IDs 179.B-2002, 195.B-0283, and 198.B-2004.
F.N.-L. gratefully acknowledges the sponsorship provided by the European Southern Observatory through a research fellowship, as well as support from the Ramón y Cajal programme (RYC2023-044924-I), funded by MCIN/AEI/10.13039/501100011033 and FSE+; grant PID2024-162148NA-I00, funded by MCIN/AEI/10.13039/501100011033 and the European Regional Development Fund (ERDF) “A way of making Europe”; and the Severo Ochoa grant CEX2021-001131-S, funded by MCIN/AEI/10.13039/501100011033. R.S. and A.M.-A. acknowledge financial support from the Severo Ochoa grant CEX2021-001131-S, funded by MCIN/AEI/10.13039/501100011033; from grant EUR2022-134031, funded by MCIN/AEI/10.13039/501100011033 and the European Union NextGenerationEU/PRTR; and from grant PID2022-136640NB-C21, funded by MCIN/AEI/10.13039/501100011033 and the European Union.
I.J.S. acknowledges financial support from the European Research Council under the ERC Consolidator Grant OPENS, GA No. 101125858, funded by the European Union.
I.J.S., L.C., and V.M.R. also acknowledge funding from grant PID2022-136814NB-I00, funded by the Spanish Ministry of Science, Innovation and Universities / State Agency of Research MICIU / AEI / 10.13039/501100011033 and by “ERDF/EU”.
R.S.K. acknowledges financial support from the ERC via Synergy Grant ECOGAL'' (project ID 855130), from the German Excellence Strategy via the Heidelberg Cluster STRUCTURES'' (EXC 2181 – 390900948), and from the German Ministry for Economic Affairs and Climate Action in project MAINN'' (funding ID 50OO2206). R.S.K. also thanks the 2024/25 Class of Radcliffe Fellows for highly interesting and stimulating discussions. Á.S.-M. acknowledges support from the RyC2021-032892-I grant funded by MCIN/AEI/10.13039/501100011033 and by the European Union ‘Next GenerationEU’ / PRTR, as well as the program Unidad de Excelencia María de Maeztu CEX2020-001058-M, and support from the PID2023-146675NB-I00 (MCI-AEI-FEDER, UE). M.C.S. acknowledges financial support from the European Research Council under the ERC Starting Grant GalFlow'' (grant 101116226) and from Fondazione Cariplo under the grant ERC attrattività n. 2023-3014.
A.G. acknowledges support from the NSF under CAREER 2142300 and AAG 2206511. COOL Research DAO \citep{2025arXiv250113160C} is a Decentralised Autonomous Organisation supporting research in astrophysics aimed at uncovering our cosmic origins. The authors thank the anonymous referee for the valuable comments.

\end{acknowledgements}

\bibliography{BibGC.bib}

\onecolumn
\appendix

\section{Tables}

\setlength{\tabcolsep}{3.5pt}
\def\arraystretch{1.2}

\begin{table*}[h]

\caption{Characterisation of the GMM decomposition of the proper motion distribution.}
\label{GMM} 
\begin{center}

\begin{tabular}{ccccccc}
 &  &  &  &  &  & \tabularnewline
\hline 
\hline 
 & $W_l$ & $\mu_l$ & $\sigma_l$ & $W_b$ & $\mu_b$ & $\sigma_b$\tabularnewline
 & (norm. units) & mas/yr & mas/yr & (norm. units) & mas/yr & mas/yr\tabularnewline
\hline 
50 km/s & 0.47 $\pm$ 0.08 & 2.99 $\pm$ 0.17 & 2.99 $\pm$ 0.22 & 0.61 $\pm$ 0.07 & 0.02 $\pm$ 0.18 & 1.46 $\pm$ 0.88\tabularnewline
 & 0.53 $\pm$ 0.08 & 1.39 $\pm$ 0.33 & 3.48 $\pm$ 0.27 & 0.39 $\pm$ 0.07 & 0.49 $\pm$ 0.30 & 4.41 $\pm$ 1.24\tabularnewline
\hline 
20 km/s & 0.68 $\pm$ 0.03 & 2.72 $\pm$ 0.06 & 1.77 $\pm$ 0.06 & 0.63 $\pm$ 0.03 & 0.10 $\pm$ 0.06 & 1.30 $\pm$ 0.06\tabularnewline
 & 0.32 $\pm$ 0.03 & 0.99 $\pm$ 0.21 & 3.98 $\pm$ 0.12 & 0.37 $\pm$ 0.03 & 0.19 $\pm$ 0.21 & 3.58 $\pm$ 0.12\tabularnewline
\hline 
Ridge & 0.71 $\pm$ 0.05 & 1.96 $\pm$ 0.14 & 2.55 $\pm$ 0.14 & 0.72 $\pm$ 0.04 & -0.11 $\pm$ 0.07 & 1.88 $\pm$ 0.43\tabularnewline
 & 0.29 $\pm$ 0.05 & 0.39 $\pm$ 0.25 & 4.77 $\pm$ 0.28 & 0.29 $\pm$ 0.04 & -0.07 $\pm$ 0.15 & 4.74 $\pm$ 0.50\tabularnewline
\hline 
Control & 0.48 $\pm$ 0.03 & 1.83 $\pm$ 0.23 & 2.05 $\pm$ 0.47 & 0.70 $\pm$ 0.03 & 0.06 $\pm$ 0.04 & 1.76 $\pm$ 0.56\tabularnewline
 & 0.23 $\pm$ 0.03 & -0.03 $\pm$ 0.32 & 4.82 $\pm$ 1.12 & 0.30 $\pm$ 0.03 & 0.00 $\pm$ 0.05 & 4.22 $\pm$ 0.51\tabularnewline
 & 0.30 $\pm$ 0.02 & -1.86 $\pm$ 0.34 & 1.94 $\pm$ 0.56 & - & - & -\tabularnewline
\hline 
 &  &  &  &  &  & \tabularnewline
\end{tabular}

\end{center}
\footnotesize
\textbf{Notes.} $W_i$, $\mu_i$, and $\sigma_i$ are the results of the GMM analysis, where $W$ represents the relative weight of each Gaussian component, and $l$ and $b$ correspond to the proper motion components parallel and perpendicular to the Galactic plane, respectively.

\end{table*}

\setlength{\tabcolsep}{5.5pt}
\def\arraystretch{1.2}

\begin{table}[h]

\caption{Dependence of $\mu_l$ on colour $H-K_s$.}

\label{table2} 
\begin{center}

\begin{tabular}{cc|cc|cc|cc}
 & \multicolumn{1}{c}{} &  & \multicolumn{1}{c}{} &  & \multicolumn{1}{c}{} &  & \tabularnewline
\hline 
\hline 
$(H-K_s)_{50}$ & $\mu_{l\ 50}$ & $(H-K_s)_{20}$ & $\mu_{l\ 20}$ & $(H-K_s)_{ridge}$ & $\mu_{l\ ridge}$ & $(H-K_s)_{control}$ & $\mu_{l\ control}$\tabularnewline
mag & mas/yr & mag & mas/yr & mag & mas/yr & mag & mas/yr\_l\tabularnewline
\hline 
1.52 $\pm$ 0.01 & 2.75 $\pm$ 0.45 & 1.51 $\pm$ 0.01 & 2.88 $\pm$ 0.24 & 1.50 $\pm$ 0.01 & 3.25 $\pm$ 0.39 & 1.494 $\pm$ 0.002 & 2.45 $\pm$ 0.15\tabularnewline
1.67 $\pm$ 0.01 & 2.50 $\pm$ 0.25 & 1.72 $\pm$ 0.01 & 2.55 $\pm$ 0.14 & 1.69 $\pm$ 0.01 & 2.95 $\pm$ 0.19 & 1.721 $\pm$ 0.002 & 1.75 $\pm$ 0.09\tabularnewline
1.89 $\pm$ 0.02 & 2.56 $\pm$ 0.35 & 1.92 $\pm$ 0.01 & 2.47 $\pm$ 0.17 & 1.92 $\pm$ 0.01 & 2.29 $\pm$ 0.26 & 1.949 $\pm$ 0.002 & 0.57 $\pm$ 0.08\tabularnewline
2.13 $\pm$ 0.03 & 3.21 $\pm$ 1.00 & 2.19 $\pm$ 0.01 & 2.03 $\pm$ 0.26 & 2.22 $\pm$ 0.01 & 0.98 $\pm$ 0.55 & 2.184 $\pm$ 0.002 & -0.25 $\pm$ 0.09\tabularnewline
- &  & - &  & - &  & 2.431 $\pm$ 0.003 & -0.54 $\pm$ 0.13\tabularnewline
- &  & - &  & - &  & 2.679 $\pm$ 0.005 & -1.11 $\pm$ 0.18\tabularnewline
\hline 
 & \multicolumn{1}{c}{} &  & \multicolumn{1}{c}{} &  & \multicolumn{1}{c}{} &  & \tabularnewline
\end{tabular}

\end{center}
\footnotesize
\textbf{Notes.} $H-K_s$ and $\mu_l$ median values obtained for the 50 and 20\,km/s clouds regions, the ridge region connecting them, and the control field.
\end{table}

\end{document}